# MinXSS-1 CubeSat On-Orbit Pointing and Power Performance
## The First Flight of the Blue Canyon Technologies XACT 3-axis Attitude Determination and Control System

James Paul Mason[1,2], Matt Baumgart[3], Bryan Rogler[3], Chloe Downs[4], Margaret Williams[1,2], Thomas N. Woods[1], Scott Palo[2], Phillip C. Chamberlin[5], Stanley Solomon[6], Andrew Jones[1], Xinlin Li[1,2], Rick Kohnert[1], and Amir Caspi[7]

[1]Laboratory for Atmospheric and Space Physics, University of Colorado, Boulder, CO, USA
[2]Aerospace Engineering Sciences, University of Colorado, Boulder, CO, USA
[3]Blue Canyon Technologies, Boulder, CO, USA
[4]University of Virginia, Charlottesville, VA, USA
[5]NASA Goddard Space Flight Center, Greenbelt, MD, USA
[6]High Altitude Observatory, Boulder, CO, USA
[7]Southwest Research Institute, Boulder, CO, USA
james.p.mason@nasa.gov

## ABSTRACT

The Miniature X-ray Solar Spectrometer (MinXSS) is a 3 Unit (3U) CubeSat designed for a 3-month mission to study solar soft X-ray spectral irradiance. The first of the two flight models was deployed from the International Space Station in 2016 May and operated for one year before its natural deorbiting. This was the first flight of the Blue Canyon Technologies XACT 3-axis attitude determination and control system – a commercially available, high-precision pointing system. We characterized the performance of the pointing system on orbit including performance at low altitudes where drag torque builds up. We found that the pointing accuracy was 0.0042° - 0.0117° (15″ - 42″, $3\sigma$, axis dependent) consistently from 190 km - 410 km, slightly better than the specification sheet states. Peak-to-peak jitter was estimated to be 0.0073° (10 s$^{-1}$) - 0.0183° (10 s$^{-1}$) (26″ (10 s$^{-1}$) - 66″ (10 s$^{-1}$), $3\sigma$). The system was capable of dumping momentum until an altitude of 185 km. We found small amounts of sensor degradation in the star tracker and coarse sun sensor. Our mission profile did not require high-agility maneuvers so we are unable to characterize this metric. Without a GPS receiver, it was necessary to periodically upload ephemeris information to update the orbit propagation model and maintain pointing. At 400 km, these uploads were required once every other week. At ∼270 km, they were required every day. We also characterized the power performance of our electric power system, which includes a novel pseudo-peak power tracker – a resistor that limited the current draw from the battery on the solar panels. With 19 30% efficient solar cells and an 8 W system load, the power balance had 65% of margin on orbit. We present several recommendations to other CubeSat programs throughout.

## 1. Introduction

The Miniature X-ray Solar Spectrometer (MinXSS, [1]) is a 3U CubeSat designed, built, and operated at the University of Colorado, Boulder (CU) Laboratory for Atmospheric and Space Physics (LASP). Its primary science objective is to measure the soft x-ray energy distribution from the sun. In order to do so, 3-axis pointing control authority and knowledge is required. For MinXSS, this capability is provided by the Blue Canyon Technologies (BCT) XACT. Two MinXSS spacecrafts were built. The second will launch in late 2017 into a ∼500 km, sun-synchronous orbit. The first was launched to the International Space Station (ISS) on 2015 December 6 and deployed from the airlock at 411 km on 2016 May 16. The last decoded packet from MinXSS-1 was received on 2017 May 6. Thus, the proposed comprehensive success criteria of 3 months of solar observations were more than satisfied. Details of the science, instruments, and calibration can be found in other recent publications [1, 2, 3]. Figure 1 shows the basic layout of MinXSS.

The standard XACT consists of three reaction wheels, three torque rods, a star tracker, a ∼110° full-cone field-of-view sun sensor, a three-axis magnetometer, an inertial measurement unit (IMU), and processing electronics to make for a simple software interface. In nominal operations for MinXSS, the only commands sent to the XACT are 1) current time update, 2) periodic ephemeris updates, and 3) the command to go to fine point mode. However, the MinXSS command and data handling (CDH) board can act as a bent pipe between ground commands and the XACT. This allows the operations team to additionally take advantage of the ∼70 commands that XACT accepts. The XACT generates ∼300 telemetry items at 5 Hz but it is only stored at 1 Hz onboard MinXSS. Only a few of these data items are pulled out into the MinXSS housekeeping packet for monitoring health and safety but the full attitude determination and control system (ADCS) telemetry packet can be downlinked real time or as a playback of stored data by command. MinXSS-1 is the first flight of the XACT unit.

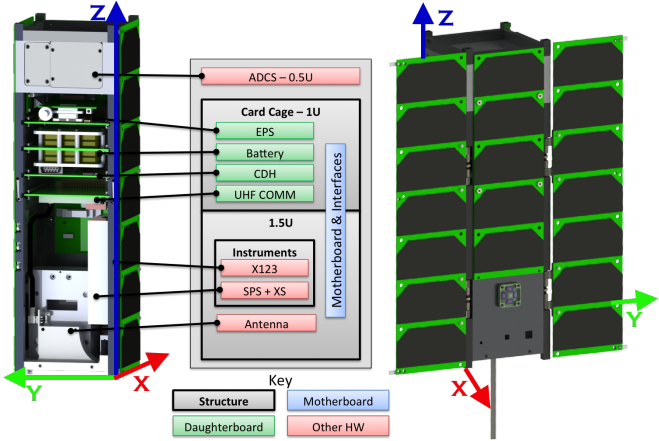

*Figure 1: MinXSS mechanical block diagram. Modified and reproduced from [1].*

The MinXSS electric power system (EPS) consists of solar panels, a battery pack, and a power routing board (herein referred to as the EPS board). There is one body-fixed solar panel with five solar cells and two deployable solar panels with seven cells each. The solar cells are AzurSpace 3G30A 30% efficient triple-junction GaAs with cover glass and tabs already installed. The full array of solar panels is capable of generating upwards of 23.41 W of power. The battery pack consists of four 2 Ah lithium-polymer batteries obtained from SparkFun (part #: 08483). They are in a 2s2p (two cells in parallel and two of those put in series) configuration to boost the voltage to 8.4 V and increase the capacity to 4 Ahr. The batteries are stacked in pairs on either side of a PCB, with a heater and temperature sensor sandwiched between each pair of batteries. Heat transfer tape was used between each layer of the pack to maximize the effectiveness of the heaters. The batteries are encapsulated with aluminum plates on top and bottom and 10 cylindrical standoffs around the edges. The battery pack underwent significant testing to satisfy the NASA human safety standards for astronauts since MinXSS-1 went to the ISS. Finally, the EPS board uses a modified direct energy transfer (DET) design that we have called pseudo-peak power tracking (PPPT). A fixed-value current limiting resistor was placed in the circuit where a maximum-PPT would normally go, and its resistance was carefully selected to prevent a power oscillation from occurring between the batteries, solar panels, and buck converters when the batteries tried to pull significant current.

The system has no shunt for excess power. Instead, excess power remains on the solar panels and manifests as heat in that location. Note that despite this, the solar panels always remained well within their temperature requirements on orbit [4]. The EPS provides the system with an unregulated battery voltage line, regulated 5 VDC line, and 3.3 VDC line. The regulated voltage lines are managed with buck converters. A separate XACT interface board was included to provide the reaction wheels with a 12 VDC line.

The spacecraft has three operational modes: Science, Safe, and Phoenix. In Science mode, everything is powered on and a housekeeping packet beacon is transmitted every N seconds, where N is configurable by command and was typically 9 to 54 for MinXSS-1. Additionally in this mode, there is ∼20 minutes total of continuous data transmission per day (which can only occur by command from the ground) and the ADCS is in "fine reference" mode. In Safe mode, the primary science instrument (called the X123) is powered off and ADCS is put into "Sun point" mode, which just uses the coarse sun sensor and IMU to find and point the solar panels toward the sun. In Phoenix mode, the X123, fine sun sensor (SPS) and x-ray photometer (XP), and ADCS are all powered off, resulting in minimum power consumption and a tumbling spacecraft. The intent of this mode is to charge the batteries in emergency conditions. The spacecraft can autonomously transition between Safe and Phoenix mode in either direction, based on a comparison between present battery voltage and a threshold specified in a table that can be changed by command. Transitions from Safe to Science can only be done by command from the ground, but the reverse can occur autonomously again based on low battery voltage.

Some aspects of pointing and power performance may have altitude dependence. As such, Figure 2 shows the altitude of MinXSS-1 over its ∼12-month mission for reference. Because the relationship between time and altitude is not linear, some plots in later sections will show a parameter versus time and altitude separately for clarity.

Section 2 will first detail the on-orbit performance of MinXSS-1 pointing. On-orbit power performance is covered in Section 3. Finally, a summary and conclusions are provided in Section 4.

## 2. On-Orbit Pointing Performance

This section characterizes pointing performance in terms of accuracy, momentum dumping, jitter, agility, orbit propagation, sensor degradation, and edge cases. Table 1 summarizes the results of the following subsections.

### 2.1. Pointing accuracy

The pointing requirements for MinXSS are 2° ($3\sigma$) accuracy and 0.05° ($3\sigma$) knowledge to meet its science goals. The specification sheet for the XACT states 0.009° ($3\sigma$) accuracy in the MinXSS X and Z axes and 0.021° ($3\sigma$) accuracy in the MinXSS Y axis. The higher precision is specified for directions across the star tracker field of view while the coarser value is along the bore sight of the star tracker. The star tracker actually points primarily in the Y-direction but is canted by 10°in the -Z direction. Thus, the specifications for the Y and Z directions are approximate values.

Table 1: MinXSS-1 pointing performance summary

| Parameter | Requirement (3σ) | Specification (3σ) | Measured Performance (3σ) |
|---|---|---|---|
| Accuracy | X: 2.0 | 0.009° | 0.0042° |
|  | Y: 2.0 | 0.021° | 0.0117° |
|  | Z: 2.0 | 0.009° | 0.0060° |
| Jitter | X: 0.3° (10 s$^{-1}$) | – | 0.0183° (10 s$^{-1}$) |
|  | Y: 0.3° (10 s$^{-1}$) | – | 0.0073° (10 s$^{-1}$) |
|  | Z: 0.3° (10 s$^{-1}$) | – | 0.0105° (10 s$^{-1}$) |
| De-tumble | – | – | <145 s |

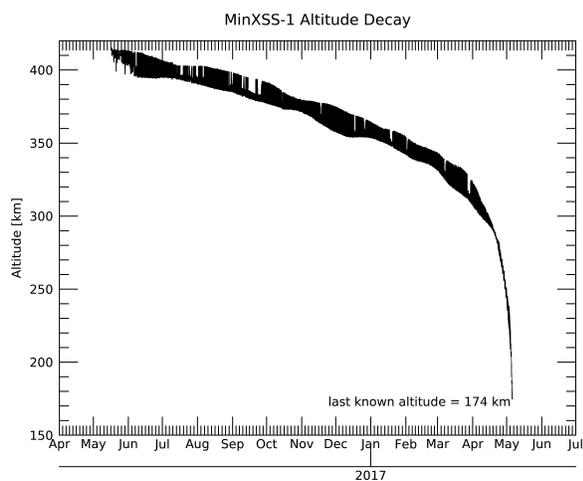

*Figure 2: MinXSS-1 altitude versus time over the ∼12-month mission.*

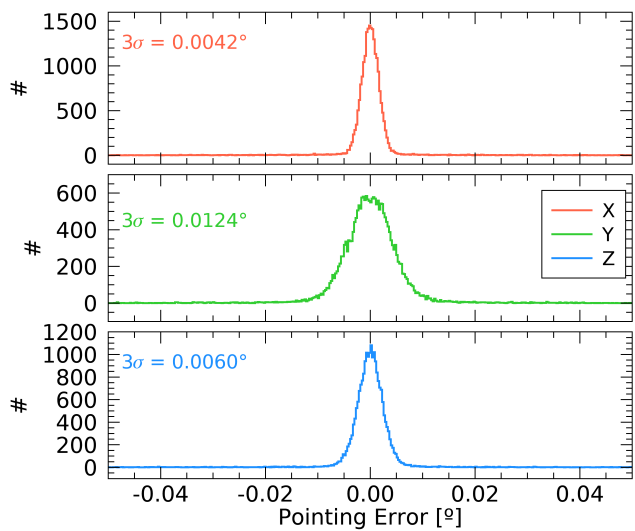

*Figure 3: Pointing accuracy histogram.*

Coarse pointing mode data were filtered out, leaving only fine reference mode attitude data, which corresponds to the MinXSS science mode. Figure 3 shows the pointing error histograms for the three axes. A full width half max was computed for each histogram. To compute the 3σ value, the full width half max was divided by a 2.355 conversion factor and then multiplied by 3. The resultant 3σ pointing accuracy for the three axes was 0.0042°, 0.0117°, and 0.0060° for X, Y, and Z, respectively. Thus the performance was ∼200-1400 times better than required by MinXSS science objectives, and ∼1.7-6 times better than the specification.

### 2.2. Momentum dumping

The implicit requirement for momentum dumping is that the torque rods must be able to shed any momentum buildup due to external torques so that the wheel speeds do not run out of momentum storage capacity during the mission. MinXSS-1 experienced increasing atmospheric drag as its orbital altitude decreased with time, starting at 411 km upon deployment from the ISS and ending at 174 km during the last contact with the spacecraft. Five times per second the XACT computes system momentum in each axis, torque rod duty cycle, and wheel speed. These values can be plotted as a function of time to determine if the system is meeting the requirement to dump momentum (Figures 4 - 6). Note that in late 2016 June, commands were sent to MinXSS to bias the system momentum in order to keep wheel speeds away from zero. These data can also be plotted as a function of altitude to easily check for any altitude-dependent trends (Figures 4 - 6). In both cases, no obvious upward trends are visible over the majority of the mission. The apparent reduction in wheel speeds until an altitude of about 250 km is an artifact of the density of downlinked data available; comparatively little time was spent at these lower altitudes so we have much less data available to plot. However, the total system momentum shown in Figure 4 shows a string of large values at about 185 km. At this altitude, the spacecraft reached the wheel cutoff threshold and the wheels were idled. At approximately 180 km, the spacecraft went into Phoenix mode, which powers off the ADCS to conserve power.

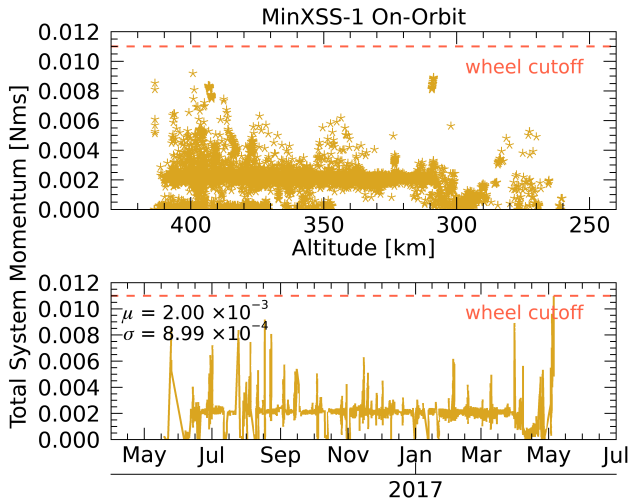

*Figure 4: Total system momentum versus time and altitude.*

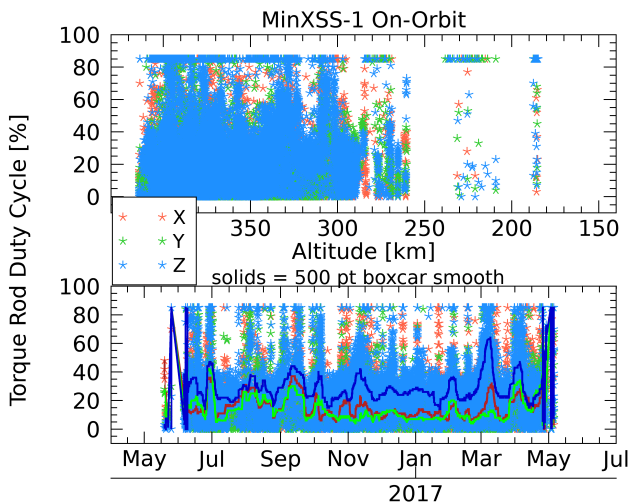

*Figure 5: Torque rod duty cycle versus time and altitude. No long term trends are visible.*

### 2.3. Jitter

The requirement on jitter was flowed down from the science objectives. With an instrument field of view of 3°

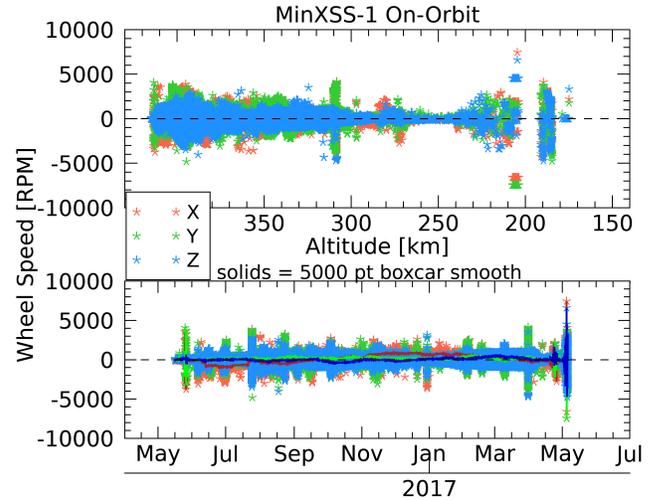

*Figure 6: Reaction wheel speeds versus time and altitude.*

and integration time of 10 seconds, the jitter in the Y and Z axes (perpendicular to the sun) must be less than 0.3° (10 s$^{-1}$) (3$\sigma$).

After filtering for just the fine reference mode data, the pointing accuracy data were placed into 10 second bins where the difference between the maximum and minimum value was computed. This represents the maximum attitude deviation in each integration period. A histogram was then generated (Figure 7). In order to compute the 3$\sigma$ values for each axis, we found the bin where >66% of the histogram data were captured, then simply multiplied by three. The resultant 3$\sigma$ values were 0.0183° (10 s$^{-1}$), 0.0073° (10 s$^{-1}$), and 0.0105° (10 s$^{-1}$) for the X, Y, and Z axes, respectively. These values are 16-41 times better than required by MinXSS science objectives. Note that these values for jitter are larger than the accuracy values in Section 2.1 because the jitter is computed as a peak-to-peak value while the pointing error centers on zero, so this estimation for jitter should be expected to be about twice as large as the pointing error value.

### 2.4. Agility

MinXSS has no explicit requirement for agility because its nominal operational mode is to stably point at the sun with a secondary constraint to to keep the antenna parallel to the ground (which equivalently keeps the star tracker pointed to zenith). Agility can be characterized three different ways: any maneuver can be limited by rate or acceleration and will have some settling time. Flight software for MinXSS has a default peak acceleration of 1 °s$^{-2}$ and peak rotation rate of 6 °s$^{-1}$, meaning that it takes 6 s to reach peak rate during which the spacecraft will rotate 18°. Therefore, any slew <36° is called "acceleration limited" because it does reach the peak rate and any slew

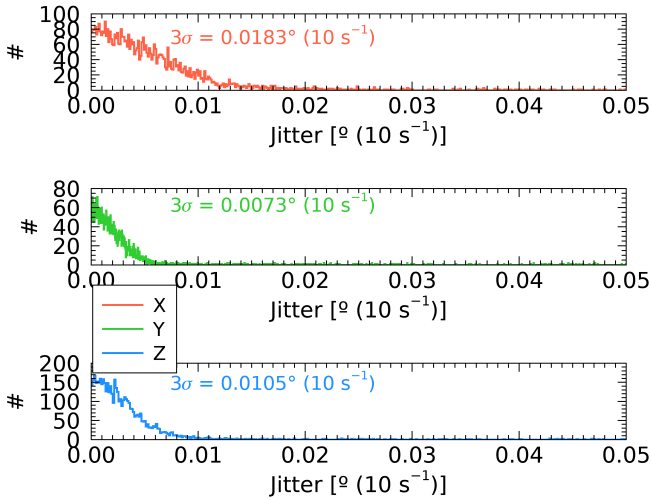

*Figure 7: Jitter histogram.*

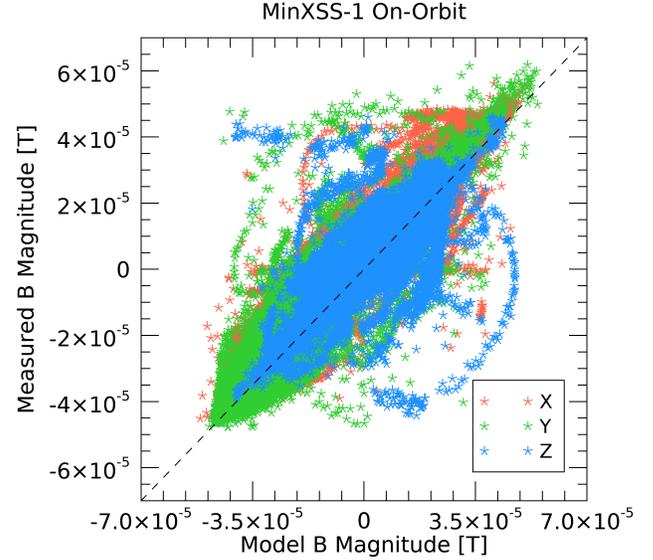

*Figure 8: Measured versus modeled magnetic field.*

>36° is called "rate limited" because the rate limit comes into play. However, the system is capable of 100 °s$^{-1}$ rates and 25 °s$^{-2}$ maneuvers and the software limits can be altered by command. MinXSS-1 did not require high-agility maneuvers so these capabilities were not tested on orbit. MinXSS-2 may perform some dedicated agility tests on orbit, at which point an addendum to this paper may be submitted.

### 2.5. Orbit propagation

MinXSS does not include the GPS unit that can be optionally provided with the XACT. As such, the spacecraft relies on an ephemeris to be uploaded for fine pointing. Near the beginning of the MinXSS-1 mission, the ephemeris only needed to be uploaded once every ∼2 weeks. This was a precautionary measure as we never performed a test to determine just how long the system could go with no new ephemeris. The onboard propagator cannot propagate forever in the presence of varying and uncertain factors such as drag. As the altitude decreased and atmospheric drag had an increasing influence on orbital position, the onboard orbit propagator built up a corresponding increase in error. At ∼350 km, the ephemeris needed to be uploaded weekly, at ∼300 km it was 3 times per week, at ∼270 km it was daily, and <∼270 km nearly every pass over the command center in Boulder would have needed to be an ephemeris upload. At that last point, the operations team chose instead to let the spacecraft fly in coarse sun pointing mode for the remainder of the mission (recall that this mode does not require ephemeris because it relies solely on the sun sensor and IMU for attitude knowledge). The error in the orbit propagator manifests in several telemetry items, which are shown in the figures of this section.

Figures 8 and 9 show the measured and modeled magnetic field. Only the model is affected by the onboard orbit propagator, so error there shows up as a discrepancy from the measured values. The few large outlier points are anomalies that are a result of the non-synchronicity of these two datasets and imperfect filtering of the data for the plot and a few times when bad ephemeris information was uploaded to the spacecraft. The sparsity of data at low altitudes is due to the operations team ceasing ephemeris uploads and data without a valid ephemeris have been filtered out of the plot. Given a good ephemeris, these data show that there is no dependence on altitude. The main impact is instead on how *long* an ephemeris remains valid. Because the operations team kept up with this until the last few weeks of the mission, the spacecraft nearly always had a valid ephemeris to propagate from. Note that uploading the ephemeris takes valuable contact time that could otherwise be used to downlink stored data. Therefore, we recommend including a GPS if possible so that ephemeris can be pulled autonomously and independently. This recommendation is strongest for CubeSats being deployed from the ISS (the majority of all CubeSats) since the low initial altitude means that extremely low altitudes are more likely to fall within the operational lifetime of the mission.

### 2.6. Sensor degradation

It is possible for the sun sensor and star tracker performance to degrade on orbit. The sun sensor itself and the neutral density filters in front of them are directly exposed to sunlight, including the particularly-damaging ultraviolet emission; and both the sun sensors and the star tracker detector are exposed to high-energy electrons and protons, all of which degrade performance [5]. One common manifestation of sensor degradation is a rise in noise over time, e.g., through energetic particle damage causing analog offsets in sensor amplifiers. Figures 10 and 11 show these sensor data. In both the sun sensor and star tracker data, the trends over time are slight. Note also that any periods of time where bright objects enter the star tracker field of view (see Section 2.7.1) may cause greater background noise.

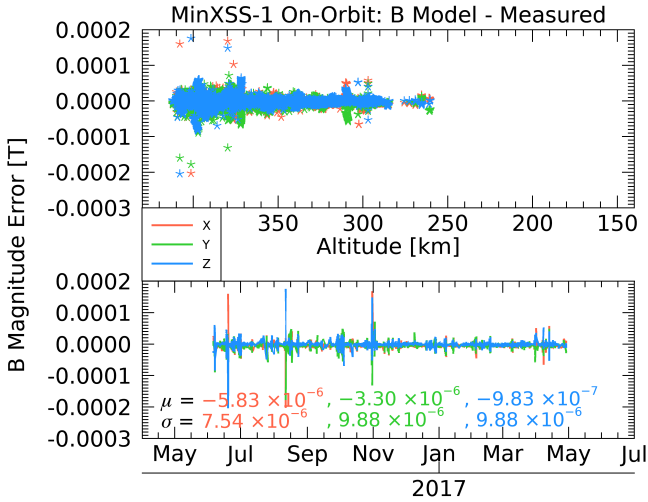

*Figure 9: Magnetic field error versus time and altitude.*

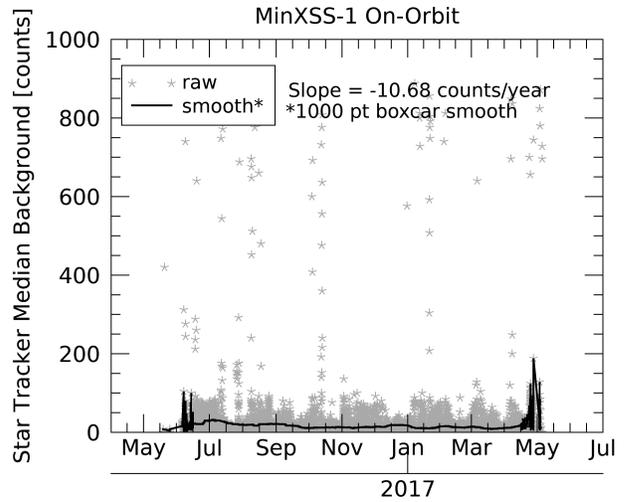

*Figure 11: Star tracker sensor median background versus time.*

gen in the atmosphere, causing them to degrade. The result was a slow increase in onboard temperatures as the spacecraft did not radiate heat as effectively and absorbed more incoming radiation [4]. As such, later in the mission the warmer temperatures may have affected the star tracker background noise. Figure 12 shows the background noise as a function of its temperature. Again, no trends are apparent with a Pearson Correlation Coefficient of only 0.18.

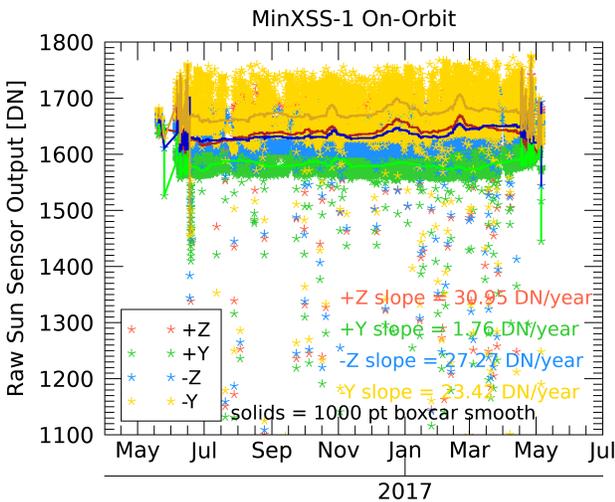

*Figure 10: Raw sun sensor output corrected for 1 AU variation versus time.*

Another source of detector noise in detectors is changes in temperature that can affect sensor electronics offset and gain. The thermal coatings on the spacecraft were exposed to ultraviolet light from the sun and oxy-

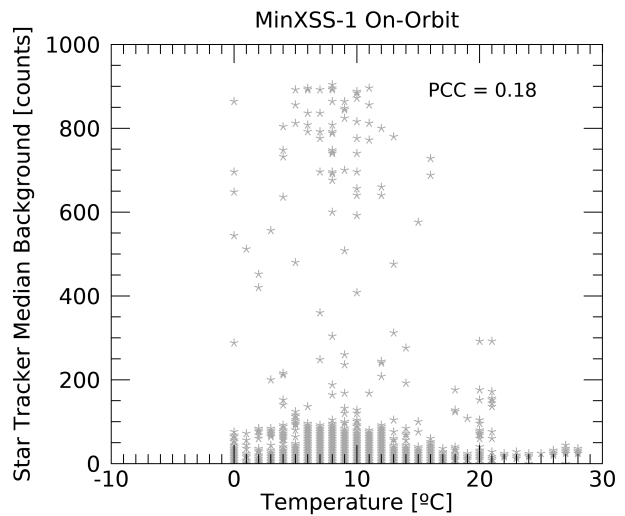

*Figure 12: Star tracker sensor median background versus temperature. The Pearson correlation coefficient is also shown.*

## 2.7. Edge cases

### 2.7.1. Celestial bodies in star tracker field of view

It is possible for bright objects to come into the field of view of the star tracker, particularly when the spacecraft is tumbling, but also in nominal operations given the pointing constraints. The primary pointing constraint for MinXSS-1 was to keep the +X axis toward the sun; the secondary constraint was to keep the +Y axis toward zenith. Note that the star tracker looks out along the +Y axis. In this configuration, the moon (either first or last quarter) occasionally passes through the star tracker field of view but the sun and earth should not. Figure 13 shows how the spacecraft performed when any of these three bodies entered the star tracker field of view. The black points are when the spacecraft is in its nominal fine reference pointing mode and the red points are when it has demoted to coarse sun pointing mode. If the celestial body prevented the ADCS from generating an attitude solution for a period of time, it would auto-demote into coarse pointing mode. As can be seen in Figure 13, the spacecraft was able to handle the moon entering the field of view of the star tracker, without falling out of fine reference mode. Note that in all of these cases, the moon was only a crescent. Plans for doing a dedicated test to intentionally put the full moon in the star tracker field of view are being worked for MinXSS-2. Also, note that instances where the sun or earth angle were low were cases when the spacecraft was either in Safe or Phoenix mode, where the ADCS was in coarse point mode or powered off completely, respectively.

### 2.7.2. De-tumble

Data captured when the spacecraft was initially deployed from the ISS were unfortunately overwritten on the onboard SD card before they could be downlinked. These data contained the initial de-tumble maneuver performed by the spacecraft. However, by the time of the first contact with our ground station, 61 minutes after deployment, the spacecraft had successfully de-tumbled. Additionally, just 11 hours before the end of the mission, another opportunity to analyze de-tumble performance arose. At this time, the spacecraft was no longer able to dump momentum fast enough to maintain pointing on the sun. As a result, the battery discharged over multiple orbits until Safe mode and then Phoenix mode were triggered. In Phoenix mode, the ADCS is powered off and the spacecraft allowed to tumble. During a subsequent ground contact with the spacecraft, the command to promote to Safe mode was sent. As part of the transition, the spacecraft turned the ADCS back on into coarse point mode, which caused it to de-tumble and try to find the sun. The operations team also commanded the spacecraft to downlink real time data generated by the ADCS during this process. The main metric of interest for de-tumbling is how long it takes. Fortunately, this command pass occurred while the spacecraft was in the sun so this timing measurement could be made. Figure 14 shows the result. Note the gap in data points during the actual maneuver. This is likely due to the null in the antenna gain passing across the ground station during the maneuver. Thus, this analysis indicates an upper-bound on the de-tumble maneuver of 145 seconds for MinXSS. The XACT is designed to complete the de-tumble maneuver within 6 minutes, assuming that the initial momentum is within the capacity of the reaction wheels and that the spacecraft is in sunlight. These results are consistent with that design. Also note that the pointing is stable once locked: the $3\sigma$ value of the displayed data is 0.0021.

## 3. On-Orbit Power Performance

This section characterizes power performance in terms of power balance, battery state of charge, and solar panel generation.

### 3.1. Power balance

Here, power balance is defined simply as

$$balance = 100 \times \frac{P_{generated} - P_{consumed}}{P_{generated}}$$

where $P$ is power and the result is expressed in percentage units. Recall that due to the PPPT EPS design, the solar panels do not always generate the maximum power they are capable of. Instead, the system only draws what it needs from the solar panels and any excess power remains on the solar cells and manifests as heat. An average power generation was computed for each spacecraft mode separately. In order to determine the total power the system is *capable* of generating on orbit, the mission data was searched for the max power generation from the solar panels (22.84 W) and use that value as $P_{generated}$ for power balance calculations. Note that in Phoenix mode, the spacecraft is tumbling so $P_{generated}$ would be some fraction of this max value, but an accounting of the actual tumble profile is beyond the scope of this paper. The final result of our analysis is shown in Table 2 and details of the procedure follow.

For each of the modes in Table 2, a specific timespan of high-cadence data while the spacecraft was in that mode was identified. Not all data generated on the spacecraft could be downlinked due to the low baud rate (9600 bits s$^{-1}$) of the communication system and limited number of passes over the ground station per day. Furthermore, instances where the spacecraft was in Phoenix mode were relatively rare. Twenty four hour periods were identified for Science and Safe modes in order to obtain an orbit average but the timespan for Phoenix mode was only 5 hours. Science and Safe mode power calculations include regular extended data downlink periods where the high-power-consuming transmitter was keyed on for several minutes at a time, which is a part of nominal operations. Phoenix mode did not include any such data playback downlinks, which is also consistent with its intended operation.

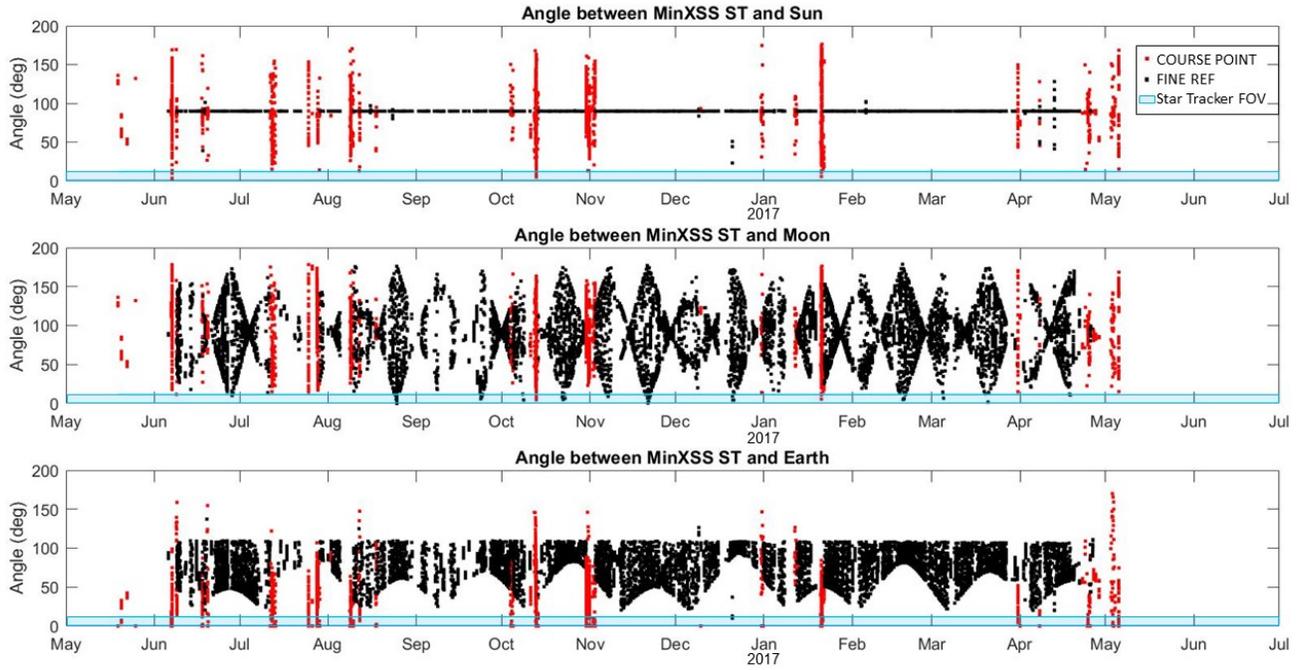

*Figure 13: Angle between star tracker bore sight and various celestial bodies versus time, with color coding for pointing mode.*

Table 2: MinXSS-1 orbit-average power

| Mode | $P_{maxgenerated}$ [W] | $P_{actualgenerated}$ [W] | $P_{consumed}$ [W] | Balance [%] |
|---|---|---|---|---|
| Science | 22.84 | 8.91 | 8.01 | 64.93 |
| Safe | 22.84 | 6.79 | 5.31 | 76.75 |
| Phoenix | 22.84 | 2.78 | 2.59 | 88.66 |

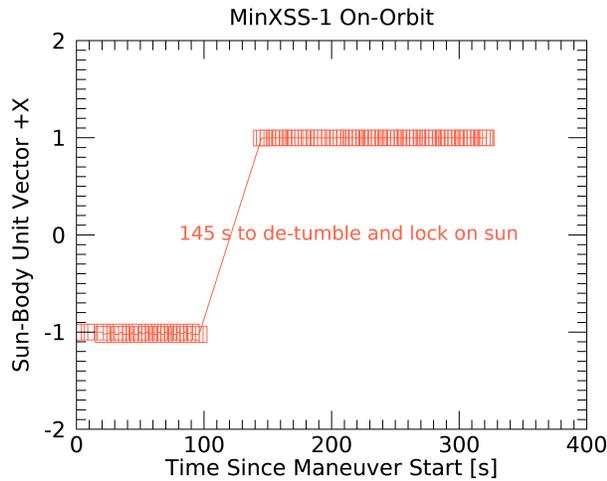

*Figure 14:* De-tumble maneuver. A +X sun-body vector of -1 indicates no signal on the sun-sensors. A value of +1 indicates perfect pointing.

During the timespan identified for each mode, power consumption and generation were calculated at each point in time and then the mean was calculated. Power consumption was computed differently dependent upon whether the spacecraft was in eclipse or sunlight. If in eclipse, the battery voltage and discharge current were multiplied. If in sunlight, the battery voltage (i.e., bus voltage) and system current draw were multiplied. For power generation, the voltage and current measured at the EPS board input from each solar panel were multiplied, then the sum taken across the three solar panels.

To conclude, the system is power positive in every mode. This does not come as a surprise because testing was done with a solar array simulator plugged in to the EPS board of the flight system for mission simulation tests on the bench and during thermal vacuum testing. A variety of orbit scenarios were simulated to ensure the system would be power positive on orbit. The first of these tests revealed that MinXSS was power negative. The design at this time used a DET EPS. The PPPT design was created out of necessity to get MinXSS in a power positive state. All subsequent mission simulation tests confirmed that the new EPS design was sufficient to keep the battery charged even in worst case conditions. The orbit performance is consistent with the tests.

### 3.2. Battery state of charge

The battery state of charge (SoC) is computed onboard by a Maxim Integrated MAX17049 fuel gauge chip that uses a proprietary algorithm. The same chip also provides a measurement of battery voltage. Several other battery voltage measurements are made throughout the MinXSS electrical system, but the fuel gauge is the most direct because it does not have any other electrical components between it and the battery. Figure 15 shows both the battery SoC and voltage as provided by the fuel gauge.

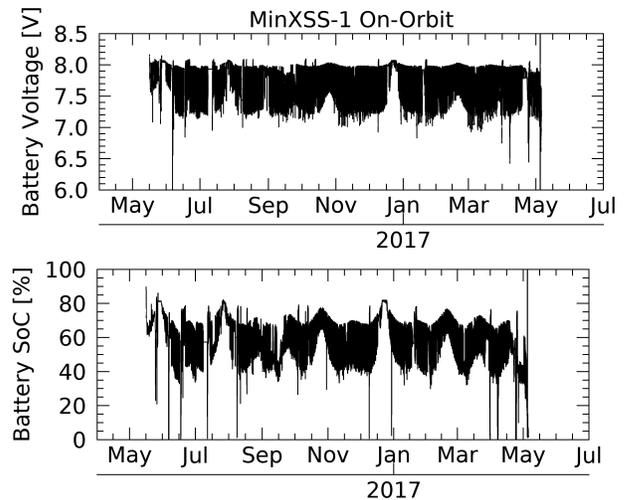

*Figure 15:* Battery voltage and state of charge as computed onboard by fuel gauge chip. Note that full charge corresponds to ∼8.5 V and at ∼6.0 V the battery protection circuits cuts the battery off from loads.

Here the result of the PPPT manifests as the battery SoC oscillating between ∼40-80% and the voltage between ∼7.2-8.0 V. The simple fixed-value current limiting resistor that is the PPPT is not capable of keeping the battery between the ideal SoC range of 80-100%, which was known from ground testing. The resistance of the PPPT was specifically selected to prevent the oscillation that occurs between a low SoC battery, which draws high current, solar panels, and buck converters. Using the DET design that was initially implemented on MinXSS resulted in approximately 50% power throughput from the solar panels to the system, which was ultimately a power negative design. With the PPPT, the system is power positive. The drawback of this system is the reduced SoC range of the battery, which negatively impacts longevity. If the PPPT had been included in the original design, then the negative consequences could have been compensated for, allowing for a higher SoC during the mission. Nevertheless, as these data clearly show, the system continued to perform for a year and it was the altitude decay rather than the battery that limited the life of the mission.

### 3.3. Solar panel power generation

Power generated by the solar panels can be simply computed by multiplying the onboard measurements of voltage and current for each panel and then summing the results. Recall that the EPS design on MinXSS results in a power generation that is less than the maximum possible. Figure 16 shows the power generation over the mission.

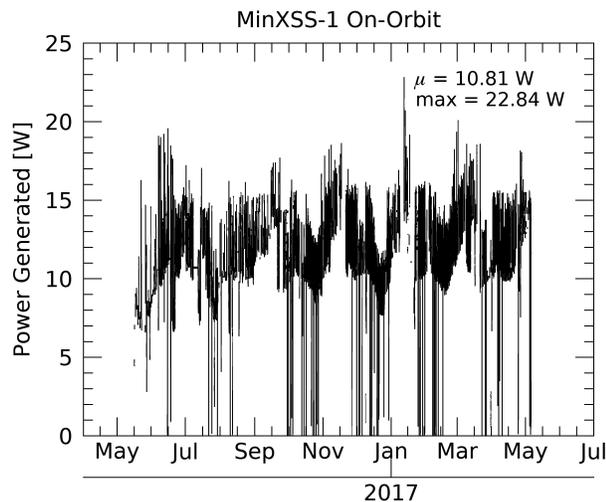

*Figure 16: Power generation from the three solar panels (19 cells total). Eclipse periods have been excluded.*

The mean power generation over the mission (excluding eclipse periods) was 10.81 W. This indicates that, on average, $< 11$ W were required to power the system and charge the battery for eclipse. Thus, it would be possible to design a system with fewer solar cells or use less efficient ones. However, Figure 16 also shows that peak power draw occasionally reached or exceeded 20 W, near the theoretical maximum (23.41 W) for the solar panels. Ultimately, this shows that the system design satisfied the actual performance requirements on orbit.

### 4. Summary and Conclusion

In each metric, the performance of the XACT exceeded the requirements flowed down from the MinXSS science objectives. The novel power system maintained a power positive balance on orbit throughout the duration of the mission in all three spacecraft modes.

We have several recommendations for other CubeSat programs that result from lessons learned and those applied for MinXSS-1. First, we strongly recommend doing functional testing of the ADCS on the ground using an air bearing table and a heliostat for sun pointing tests, as was done for MinXSS. This ensures that the ADCS will point the right face of the spacecraft toward the sun, which can identify any small but mission-ending ADCS sensor or actuator issues or phasing mistakes such as from mounting sun sensors in the wrong orientation, and negative signs in the wrong place in flight code. Next, if going into a low altitude orbit, either include a GPS for onboard autonomous ephemeris updates or include routine ephemeris uploads in the operations plan (increasing in frequency as altitude comes down). Also, if including a star tracker, run an orbit simulation to see if bright objects will enter the FOV. If so, run analysis and tests to ensure the star tracker can handle these situations. Identify this as a risk and mitigate appropriately. Include an autonomous emergency power mode, whose entire purpose is to charge the battery. Finally, test the full spacecraft with a solar array simulator and a variety of insolated/eclipse periods to ensure the actual system is power positive in all conditions.

### Acknowledgments

The authors would like to thank the entire 80+ person MinXSS CubeSat team for their contributions to this successful program. This work was supported by NASA grant NNX14AN84G.